\documentclass[seceq]{ptptex}
\usepackage{amsmath,amssymb,color,graphics,amscd,amsfonts,epsf}

\title{
Curved Superspaces and Local Supersymmetry\\
 in Supermatrix Model
}

\author{
Masanori \textsc{Hanada}$^{1}$, Hikaru \textsc{Kawai}$^{1,2}$ 
and Yusuke \textsc{Kimura}$^{1}$
}

\inst{
${}^1$ {\it Department of Physics, Kyoto University, 
Kyoto 606-8502, Japan} \\
${}^2$ {\it Theoretical Physics Laboratory, RIKEN, Wako 351-0198, 
Japan}
\\
\vspace{1cm}
{\sf hana,hkawai,ykimura@gauge.scphys.kyoto-u.ac.jp}
}
\vspace{1cm}

\abst{
In a previous paper, we introduced a new interpretation 
of matrix models, in which any $d$-dimensional curved space 
can be realized in terms of $d$ matrices, and 
the diffeomorphism and the local Lorentz symmetries are included 
in the ordinary unitary symmetry of the matrix model. 
Furthermore, we showed that the Einstein equation is 
naturally obtained, 
if we employ the standard form of the action, 
$S=-tr\left([A_a,A_b][A^a,A^b]\right)+\cdots $. 
In this paper, we extend this formalism to include supergravity.
We show that the supercovariant derivatives on any 
$d$-dimensional curved space can be expressed in 
terms of $d$ supermatrices, and the local supersymmetry 
can be regarded as a part of the superunitary symmetry.
We further show that the Einstein and Rarita-Schwinger equations
are compatible with the supermatrix generalization of the standard action. 
}

\begin{document}

\maketitle

\section{Introduction}
Although it is believed that 
string theory can provide a formalism describing 
the unification of the fundamental interactions, 
its present formulation based on perturbation theory is not satisfactory. 
In order to examine whether it actually describes our four-dimensional world,
a non-perturbative, background independent formulation is needed. 
Matrix models represent a promising approach to the study of the nonperturbative 
dynamics of string theory. 
For critical strings, 
they are basically obtained through dimensional reduction of 
the ten-dimensional $U(N)$ ${\cal N}=1$ supersymmetric Yang-Mills 
theory \cite{BFSS,IKKT}. 

IIB matrix model \cite{IKKT} 
is obtained through dimensional reduction to 
a point, and the action is given by 
\begin{eqnarray}
  S=-\frac{1}{g^2}tr
  \left(
    \frac{1}{4}[A_{a},A_{b}][A^{a},A^{b}]
    +
    \frac{1}{2}\bar{\psi}\gamma^a[A_{a},\psi]
  \right), 
  \label{action_IIB}
\end{eqnarray}
where $\psi$ is a ten-dimensional Majorana-Weyl spinor, and 
$A_{a}$ and $\psi$ are $N\times N$ Hermitian matrices. 
The indices $a$ and $b$ are 
contracted by the flat metric. 
This action has an $SO(10)$ global Lorentz symmetry and $U(N)$ symmetry. 
A problem with this model is that 
it is unclear how curved spaces are described and 
how the fundamental principle of general relativity 
is realized in it. 

It may seem that the space of dynamical variables 
would become very small through the dimensional reduction, 
but this is not the case 
if $N$ is infinitely large \cite{EK}. 
Indeed, IIB matrix model contains 
super Yang-Mills theory:
\begin{eqnarray}
  \mbox{$10D$ super Yang-Mills}
  \overset{{\rm dim.red.}}{\longrightarrow}
  \mbox{IIB matrix model}
  \overset{N\to\infty}{\supset}
  \mbox{$10D$ super Yang-Mills} . 
  \nonumber
\end{eqnarray}
The matrix-valued variables $A_{a}$ and $\psi_{\alpha}$ act on 
the Hilbert space $V={\mathbb C}^N$ as endomorphisms, i.e. 
linear maps from $V$ to itself. 
Because $V$ is infinite dimensional 
in the large-$N$ limit, we can interpret it in  various ways. 
If we assume that $V$ is the space of an $n$-component complex scalar field, 
$V=\{\varphi^i:{\mathbb R}^{10}\to{\mathbb C}^n\}$, 
instead of $V={\mathbb C}^N$, 
then an endomorphism $T$ is a bilocal field 
$K^{ij}(x,y)$, which can be formally regarded as 
a set of differential operators of arbitrary rank 
with $n\times n$ matrix coefficients \cite{hep-th/0204078}:
\begin{eqnarray}
  (T\varphi)^i(x)
  &=&
  \sum_{j=1}^n\int d^{10}y K^{ij}(x,y)\varphi^j(y)
  \nonumber\\
  &=&
  c_{(0)}^{ij}(x)\varphi^j(x)
  +
  c_{(1)}^{\mu,ij}(x)\partial_\mu\varphi^j(x)
  +
  c_{(2)}^{\mu\nu,ij}(x)\partial_\mu\partial_\nu\varphi^j(x)
  +
  \cdots. 
\end{eqnarray}
In particular, we can consider the covariant derivative 
as a special value of $A_a$: 
\begin{eqnarray}
  A_a
  =i\left( \partial_a-ia_a(x) \right)
  \in End(V). 
\end{eqnarray}
In this sense, the ten-dimensional $U(n)$ ${\cal N}=1$ 
super Yang-Mills theory can be embedded 
in the matrix model, and local gauge symmetry is realized 
as a part of the $U(N)$ symmetry of the matrix model; 
that is, we have  
\begin{eqnarray}
  \delta A_a=i[\lambda,A_a], 
\end{eqnarray}
where $\lambda$ is a matrix-valued function on ${\mathbb R}^{10}$, 
which is a $0$-th order differential operator in $End(V)$. 
This interpretation of $V$, however, 
does not manifest the existence of gravity. 
Therefore it is desirable to formulate another interpretation in which 
the diffeomorphism and the local Lorentz symmetries are embedded 
in $U(N)$. 
With such an interpretation, 
any curved space corresponds to a certain matrix configuration, 
and the path integral includes 
the summation of all the curved spaces.  
Our idea is that it is not a curved space itself 
but, rather, the covariant derivatives on it that correspond to matrices. 
Note that all the information needed to 
describe the physics on a manifold are contained 
in the covariant derivatives. 

Suppose we have a curved space $M$ with a fixed spin structure 
and a covariant derivative on it\footnote{
Because a covariant derivative introduces a new vector index, 
it is  not an endomorphism. Therefore, 
it cannot be represented by a set of matrices. 
} . Our task is to find 
\begin{eqnarray}
&&\hspace{0cm}(a)\quad \mbox{a good space $V$ and} \cr 
&&\hspace{0cm}(b)\quad 
\mbox{a good object $\nabla_{(a)}$ that is equivalent to 
a covariant derivative $\nabla_a$ }
\nonumber
\end{eqnarray}
such that each component of $\nabla_{(a)}\ (a=1,2,\cdots,10)$ 
is expressed as an endomorphism on $V$. 
In a previous paper \cite{HKK}, 
we showed that $V$ is given by the space 
of functions on the principal $Spin(10)$ bundle over $M$ 
and that $\nabla_{(a)}$ is given by 
\begin{equation}
\nabla_{(a)}=R_{(a)}{}^b(g^{-1})\nabla_b, 
\end{equation}
where $\nabla_{a}$ is the covariant derivative 
\begin{equation}
  \nabla_a
  =e_a{}^{m}(x)
  \left(
    \partial_m-\omega_m{}^{bc}(x){\cal O}_{bc}
  \right), 
\end{equation}
and $R_{(a)}{}^{b}(g)$ is the vector representation of $Spin(10)$. 
Here we assume that all indices are Lorentz. 
We can show that each component of $\nabla_{(a)}$ is indeed 
an endomorphism on $V$, while $\nabla_{a}$ is not. 
In \S \ref{sec:Describing Curved Spaces by Matrices}, 
we review this formulation in detail. 
By introducing the space $V$ given above,  
we can successfully embed gravity in the matrix model  
in such a way that the diffeomorphism and the local Lorentz 
symmetry are realized as parts of the $U(N)$ symmetry of 
the matrix model. 

In contrast to the situation with gravity, however, 
supergravity cannot be easily embedded in the matrix model 
in this interpretation,  
because the $U(N)$ symmetry does not contain 
a fermionic symmetry.  
In order to implement a manifest local 
supersymmetry in the matrix model, 
we need to extend the space $V$ and the $U(N)$ symmetry to include 
a fermionic parameter.  
This is done by 
extending $M$ to a curved superspace and taking $V$ to be 
the space of functions on the principal $Spin(10)$ bundle over 
the superspace. 
This extension forces us to consider supermatrices instead of 
ordinary matrices. 

The organization of this paper is as follows. 
In the next section, we review our previous paper \cite{HKK}. 
First we explain how a covariant derivative 
on a $d$-dimensional 
curved space can be expressed as a set of $d$ endomorphisms. 
Then we introduce a new interpretation 
in which matrices represent such differential operators. 
Based on this interpretation, 
we can show that the Einstein equation 
follows from the equation of motion of the matrix model. 
In \S \ref{sec:Differential Operators on Superspace}, 
we express a supercovariant derivative 
on a curved superspace in terms of endomorphisms.  
Although we consider only ${\cal N}=1$ supersymmetry, 
the generalization to ${\cal N}>1$ is straightforward. 
In \S \ref{sec:Supermatrix model}, 
we introduce supermatrix models. 
Dynamical variables are regarded as supercovariant derivatives on 
a curved superspace,   
and the symmetries of supergravity are 
embedded into the superunitary symmetry. 
We show that the equations of motion of the supermatrix model are satisfied 
if we assume the standard torsion constraints 
and the supergravity equations of motion.   
\S \ref{sec:Conclusions and Discussions} is devoted 
to conclusions and discussion. 
In Appendix \ref{appendix:supermatrix} we summarize the properties of 
supermatrices. In Appendices \ref{sec:Derivation of EOM} 
and \ref{subsec:kappa=0}, we present the detailed calculations 
in deriving a classical solution of the supermatrix model.  
\section{Describing curved spaces by matrices}
\label{sec:Describing Curved Spaces by Matrices}
In this section, we summarize the results given in Ref. \citen{HKK}. 
We first explain how 
a covariant derivative on a $d$-dimensional 
manifold $M$ 
can be expressed by a set of $d$ endomorphisms acting 
on the space of functions on the principal $Spin(d)$-bundle 
over $M$. 
We then apply this idea to the matrix model and 
introduce a new interpretation  
in which dynamical variables represent   
differential operators on curved spaces. 
We show that the matrix model reproduces  
the Einstein equation correctly, 
and that the symmetries of general relativity are 
realized as parts of the $U(N)$ symmetry. 
\subsection{Covariant derivative as a set of  endomorphisms}
\label{subsec:Covariant Derivative as a Set of  Endomorphisms}
Let $M$ be a Riemannian manifold with a fixed 
spin structure and $M=\cup_i U_i$ be its open covering. 
On each patch $U_i$, the covariant derivative is expressed as  
\begin{eqnarray}
  \nabla_a^{[i]}
  =e_a{}^{m[i]}(x)
  \left(
    \partial_m-\omega_m{}^{bc[i]}(x){\cal O}_{bc}
  \right), 
\end{eqnarray}
where $e_a{}^m(x)$ and $\omega_m{}^{bc}(x)$ are the vielbein 
and the spin connection, respectively. 
${\cal O}_{bc}$ is the Lorentz generator that acts on 
Lorentz indices. 
The index $[i]$ is the label of the patch. 
In the overlapping region $U_i\cap U_j$, 
the operators $\nabla_a^{[i]}$ and $\nabla_a^{[j]}$ are related by 
\begin{eqnarray}
  \nabla_a^{[i]}
  =
  R_a{}^b(t_{ij}(x))\nabla_b^{[j]}, 
\end{eqnarray}
where $t_{ij}:U_i\cap U_j\to G=Spin(d)$ is the transition function, 
and $R_a{}^b(t_{ij}(x))$ is the vector representation of $t_{ij}(x)$. 

Let us consider the principal $G$ bundle on $M$ associated 
with the spin structure and denote it by $E_{prin}$. 
It is constructed from the set $U_i \times G$ 
by identifying 
$(x_{[i]},g_{[i]})$ with $(x_{[j]},g_{[j]})$:
\begin{eqnarray}
  x_{[i]}=x_{[j]}, 
  \qquad
  g_{[i]}=t_{ij}(x)g_{[j]}.
  \label{gluing condition}
\end{eqnarray}
We take $V=C^{\infty}(E_{prin})$, which is the space of 
smooth functions on $E_{prin}$. 
We assume that covariant derivatives act on the space $V$, 
that is, ${\cal O}_{ab}$ generates an infinitesimal left action, 
\begin{eqnarray}
  i\epsilon^{ab}
  \left(
    {\cal O}_{ab}f^{[i]}
  \right)(x,g)
  =
  f^{[i]}\left(
    x, \left(1+i\epsilon^{ab}M_{ab}\right)^{-1}g
  \right)
  -
  f^{[i]}\left(
    x,g
  \right),   
\end{eqnarray}
where $M_{ab}$ is the matrix of the fundamental representation. 
Then, 
we can construct endomorphisms from a covariant derivative 
in the following way:
\begin{eqnarray}
  \nabla_{(a)}^{[i]}=R_{(a)}{}^b(g_{[i]}^{-1})\nabla_b^{[i]}. 
\end{eqnarray}
Here, $R_{(a)}{}^b(g)$ is the vector representation of $G$.\footnote{
$R_{a}{}^b$ and $R_{(a)}{}^b$ represent the same quantity. 
However, we distinguish them, because $a$ and $(a)$ obey different 
transformation laws. 
Specifically, $a$ is transformed by the action of $G$, 
while $(a)$ is not. 
}
In the overlapping region of two patches, 
the actions of $\nabla_{(a)}^{[i]}$ and $\nabla_{(a)}^{[j]}$ 
on $f\in V=C^{\infty}(E_{prin})$ are related as 
\begin{eqnarray}
  \nabla^{[i]}_{(a)}f^{[i]}(x_{[i]},g_{[i]})
  &=&
  R_{(a)}{}^b(g_{[i]}^{-1})\nabla_b^{[i]}
  f^{[i]}(x_{[i]},g_{[i]})
  \nonumber\\
  &=&
  R_{(a)}{}^b(g_{[i]}^{-1})R_b{}^c(t_{ij}(x))\nabla_c^{[j]}
  f^{[j]}(x_{[j]},g_{[j]})
  \nonumber\\
  &=&
  R_{(a)}{}^c
  \left(
    (t_{ij}(x)^{-1}g_i)^{-1}
  \right)\nabla_c^{[j]}
  f^{[j]}(x_{[j]},g_{[j]})
  \nonumber\\
  &=&
  R_{(a)}{}^c\left(g_{[j]}^{-1}\right)
  \nabla_c^{[j]}
  f^{[j]}(x_{[j]},g_{[j]})
  \nonumber\\
  &=&
  \nabla_{(a)}^{[j]}f^{[j]}(x_{[j]},g_{[j]}),  
\end{eqnarray}
where $f^{[i]}(x_{[i]},g_{[i]})$ is the expression of $f$ 
on $U_i\times G$. Note that with the identification 
given in (\ref{gluing condition}), we have 
$f^{[i]}(x_{[i]},g_{[i]})=f^{[j]}(x_{[j]},g_{[j]})$, by definition. 
This confirms that each component of $\nabla_{(a)}$ is {\it globally} 
defined on $E_{prin}$ and is indeed an endomorphism on $V$. 
The index $(a)$ merely labels $d$ endomorphisms. 
Similarly, operators with vector or spinor indices  
can be mapped to a set of endomorphisms using  
the vector and spinor representation matrices of $G$, 
for example, 
\begin{eqnarray}
  T_{(\alpha)}
  &=&
  R_{(\alpha)}{}^\beta(g^{-1}) T_\beta, 
  \nonumber\\
  T_{(a)(b)}
  &=&
  R_{(a)}{}^{c}(g^{-1}) 
  R_{(b)}{}^{d}(g^{-1})T_{cd}.    
\end{eqnarray}
Here $R_{(a)}{}^b(g^{-1})$ and $R_{(\alpha)}{}^\beta(g^{-1})$ 
are the vector and spinor representations, respectively. 
In addition, we have the relation  
\begin{eqnarray}
  \nabla_{(a)}f_{b}(x,g)
  =
  R_b{}^{(c)}(g)\nabla_{(a)}f_{(c)}(x,g),
  \label{extension of covariant derivative}
\end{eqnarray}
where the covariant derivative on the left-hand side 
also acts on the vector index $b$. 
This point is explained in detail in Ref. \citen{HKK}.  

The method presented in this subsection is valid 
in any number of dimensions, and 
we can express the covariant derivative on any $d$-dimensional 
Riemannian manifold in terms of $d$ matrices. 
\subsection{A new interpretation of IIB matrix model}
\label{subsec:New interpretation of IIB matrix model}
In this subsection we present a new interpretation of 
IIB matrix model \cite{HKK}. 
As we showed in the previous subsection, 
for any $d$-dimensional Riemannian manifold with a fixed 
spin structure, its covariant derivative can be described 
by a set of $d$ matrices acting on $C^\infty(E_{prin})$. 

\begin{figure}[htb]
  \begin{center}
    \scalebox{0.40}{
      \includegraphics[25cm,16cm]{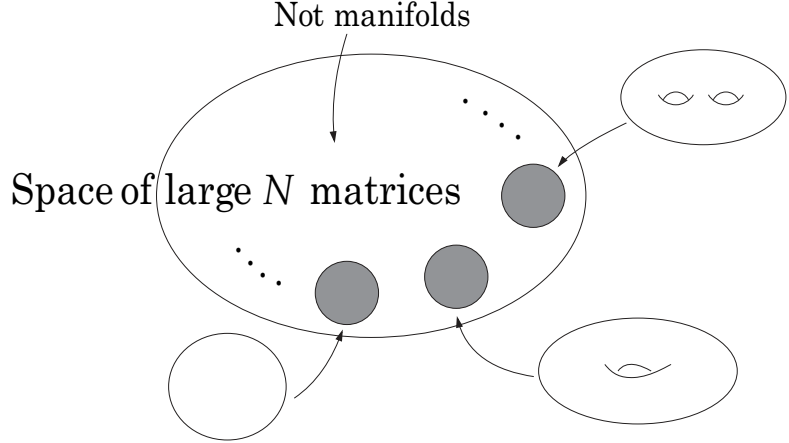}
    }
    \caption{
      Path integral include all the curved spaces. 
    }
    \label{figure}
  \end{center}
\end{figure}

Let us consider the space of large-$N$ matrices. 
For any manifold $M$, this space 
contains a set of matrices
which are unitary equivalent to 
the covariant derivatives $i\nabla_{(a)}$ on this manifold.  
If the matrices $A_a$ are sufficiently close to 
the derivatives $i\nabla_{(a)}$ on one of the manifolds $M$, 
it is natural to regard these $A_a$ as acting on $C^\infty(E_{prin})$,   
and to expand $A_a$ about $i\nabla_{(a)}$:\footnote{
Strictly speaking, because $A_a$ is Hermitian, we should introduce  
the anticommutator $\{\ ,\ \}$ in Eq. 
(\ref{eq:expansion by local fields}):
\begin{eqnarray}
  A_a
  &=&
  i\nabla_{(a)}
  +
  a_{(a)}(x,g)
  +
  \frac{i}{2}\{a_{(a)}{}^{(b)}(x,g),\nabla_{(b)}\}
  +
  \frac{i}{2}\{a_{(a)}{}^{bc}(x,g),{\cal O}_{bc}\}
  +
  \cdots.   
\end{eqnarray}
In the following, we omit the anticommutator 
for simplicity. 
} 
\begin{eqnarray}
  A_a
  &=&
  i\nabla_{(a)}
  +
  a_{(a)}(x,g)
  +
  ia_{(a)}{}^{(b)}(x,g)\nabla_{(b)}
  +
  ia_{(a)}{}^{bc}(x,g){\cal O}_{bc}
  \nonumber\\
  & &
  \quad
  +
  i^2a_{(a)}{}^{(b)(c)}(x,g)\nabla_{(b)}\nabla_{(c)}
  +
  i^2a_{(a)}{}^{(b),cd}(x,g)\nabla_{(b)}{\cal O}_{cd}
  +
  \cdots. 
  \label{eq:expansion by local fields}
\end{eqnarray}
In this expansion, local fields appear as coefficients.  
For example, $a_{(a)}{}^{(b)}(x,g)$ contains  
a fluctuation of the vielbein. 
Coefficients of higher-order derivative terms 
correspond to higher-spin fields. 
In this sense, this part of the space of large-$N$ matrices 
describes the dynamics around the background spacetime $M$. 
Some matrix configurations correspond to $S^{10}$, 
others to $T^{10}$, and so on (see Fig.\ref{figure})\footnote{
A matrix configuration in which some of matrices $A_a$ fluctuate 
around $0$ corresponds to a lower-dimensional manifold. 
}.   

Note that {\it we do not fix} $M$ and that 
{\it all Riemannian manifolds with all possible 
spin structures are included in the path integral}. 
Of course, it also contains matrix configurations 
that are not related to any manifold.  
For example, the matrices $A_a$ satisfying 
$[A_{a},A_{b}]
=iB_{ab}\cdot\textbf{1}
$ with constant $B_{ab}$
describe a flat noncommutative space.  
Another trivial example is 
$A_{a}=\textbf{0}\ (a=1,2,\cdots,10)$. 

\vspace{0.5cm}
\noindent
\textbf{Classical solutions}\\

Now let us consider the equation of motion. 
The action of IIB matrix model is given by Eq. (\ref{action_IIB}).  
Variation with respect to $A_{a}$ gives the equation of motion 
\begin{eqnarray}
  \left[
    A^{a},[A_{a},A_{b}]
  \right]
  =0. 
  \label{eq:EOM with parenthesis_1}
\end{eqnarray}
Here, we simply set $\psi=0$. 
If we impose the ansatz
\begin{eqnarray}
  A_{a}=i\nabla_{(a)}, 
  \label{ansatz:bosonic case}
\end{eqnarray}
then Eq. (\ref{eq:EOM with parenthesis_1}) becomes
\begin{eqnarray}
  \left[
    \nabla^{(a)},\left[\nabla_{(a)},\nabla_{(b)}\right]
  \right]
  =0. 
  \label{eq:EOM with parenthesis_2}
\end{eqnarray}
Equation (\ref{eq:EOM with parenthesis_2}) is equivalent to\footnote{
  We use the formula (\ref{extension of covariant derivative}) and 
  ${}^T R= R^{-1}$. 
}
\begin{eqnarray}
  0
  &=&
  \left[
    \nabla^{a},\left[\nabla_{a},\nabla_{b}\right]
  \right]
  \nonumber\\
  &=&
  [\nabla^a,R_{ab}{}^{cd}{\cal O}_{cd}]
  \nonumber\\
  &=&
  \left(\nabla^a R_{ab}{}^{cd}\right){\cal O}_{cd}
  -
  R_b{}^c\nabla_c, 
  \label{eq:EOM without parenthesis}
\end{eqnarray}
where $R_{ab}{}^{cd}$ is the Riemann tensor and 
$R_b{}^c=R_{ab}{}^{ac}$ is the Ricci tensor. 
Here we have assumed that $\nabla_a$ is torsionless. 
Note that Eq. (\ref{eq:EOM without parenthesis}) 
holds if and only if we have 
\begin{eqnarray}
  \nabla^a R_{ab}{}^{cd}=0, 
  \quad
  R_{ab}=0. 
\end{eqnarray}
The first equation here follows from the second by the Bianchi identity 
$\nabla^{[a}R_{bc}{}^{de]}=0$. 
Therefore, the covariant derivative on a Ricci-flat spacetime 
is a classical solution.

We next consider a matrix model with a mass term: 
\begin{eqnarray}
  S=
  -\frac{1}{g^2}
  tr\left(
    \frac{1}{4}[A_a,A_b][A^a,A^b]
    +
    \frac{1}{2}\bar{\psi}\gamma^a[A_{a},\psi]    
  \right)
  +
  \frac{m^2}{g^2}tr\left(A_a A^a\right). 
\end{eqnarray}
The equation of motion is now given by 
\begin{eqnarray}
  [A_b,[A_a,A^b]]-2m^2A_a=0, 
\end{eqnarray}
which leads to 
\begin{eqnarray}
  R_a{}^b=-2m^2\delta_a{}^b. 
\end{eqnarray}
This is the Einstein equation with a cosmological constant 
\begin{eqnarray}
  \Lambda=-8m^2. 
\end{eqnarray}
However, if $m^2>0$, the configuration $A_a=0\ (a=1,\cdots,10)$ 
is also a classical solution and 
has lower free energy. 
Therefore, such a configuration may dominate the path integral. 
If the positive mass term exists only for $d$ directions, 
$S_{{\rm mass}}=\frac{m^2}{g^2}\sum_{a=1}^dtr\left(A_a A^a\right)$, 
then we expect that 
these $d$ directions are collapsed to a point 
and the rest $10-d$ directions remain flat. 

The analysis of the eigenvalue distribution \cite{eigenvalue} 
is consistent with this expectation. 
Suppose an effective mass term is 
generated in six directions and not in the rest four directions. 
Then that analysis indicates 
that the four dimensional spacetime is generated.  
\\
\\
\noindent
\textbf{Diffeomorphism and local Lorentz invariance}\\
\label{subsubsec:Diffeo and LL}

We now show that the symmetries of general relativity 
are realized as parts of the $U(N)$ symmetry of the matrix model. 
The $U(N)$ symmetry of the matrix model is expressed as 
\begin{eqnarray}
  \delta A_{a}
  =
  i[\Lambda,A_{a}], 
  \label{eq:unitary transformation}
\end{eqnarray}
where $\Lambda$ is a Hermitian matrix. 
If we expand $A_a$ as Eq. (\ref{eq:expansion by local fields}) and take 
\begin{eqnarray}
  \Lambda
  =
  \frac{i}{2}\{\lambda^{(a)},\nabla_{(a)}\}
  =
  \frac{i}{2}\{\lambda^{a}(x),\nabla_{a}\}, 
\end{eqnarray}
Eq. (\ref{eq:unitary transformation}) becomes a diffeomorphism on the
fields that appear in Eq. (\ref{eq:expansion by local fields}), and 
$e_a{}^m$ and $\omega_m{}^{ab}$ that appear in $\nabla_{(a)}$. 
For example, we have  
\begin{eqnarray}
  \delta e_a{}^m(x)
  &=&
  e_a{}^n(x)\nabla_n\lambda^m(x), 
  \\
  \delta\omega_m{}^{ab}(x)
  &=&
  \lambda^n(x)R_{mn}{}^{ab}(x), 
  \\
  \delta a_a(x)
  &=&
  -\lambda^n\nabla_n a_a(x). 
\end{eqnarray}
This is the transformation law of the diffeomorphism. 
Similarly, if we take 
\begin{eqnarray}
  \Lambda=i\lambda^{ab}(x){\cal O}_{ab}, 
\end{eqnarray}
we obtain 
\begin{eqnarray}
  \delta e_a{}^m(x)
  &=&
  -\lambda_a{}^b(x) e_b{}^m(x), 
  \\
  \delta\omega_m{}^{ab}(x)
  &=&
  -
  \partial_m\lambda^{ab}(x)
  +
  2\lambda^{[a}{}_c(x)\omega_m{}^{b]c}(x), 
\end{eqnarray}
which is the local Lorentz transformation. 
\section{Supercovariant derivative as a set of endomorphisms}
\label{sec:Differential Operators on Superspace}
In the last section we showed that the symmetries of 
general relativity can be embedded in the $U(N)$ symmetry 
of the matrix model. 
In order to implement local supersymmetry, 
we employ a superspace ${\cal M}$ as the base manifold  
and consider the supercovariant derivative. 
\subsection{General prescription}
\label{subsec:General Prescription}
In this subsection, we introduce the supercovariant derivative on 
a $d$-dimensional ${\cal N}=1$ superspace 
and explain how to relate it to a set of endomorphisms. 

The coordinates of the superspace 
are $z^M=(x^m,\theta^\mu)$, where $x^m\ (m=1,2,\cdots,d)$ 
are the bosonic components 
and $\theta^\mu\ (\mu=1,2,\cdots,d_s)$ are the fermionic components. 
The quantity $d_s$ is the dimension of the spinor representation. 
For example, if $d=10$, $\theta$ corresponds to 
the Majorana-Weyl spinor and $d_s=16$. 
If $d=4$, $\theta$ corresponds to the Majorana spinor and $d_s=4$. 
Also, $M=(m,\mu)$ is an Einstein index, which can be converted into 
a Lorentz index, $A=(a,\alpha)$, using the supervielbein $E_A{}^M(x,\theta)$. 
In the following, we denote the Einstein indices by $M,N,L,\cdots$ and 
the Lorentz indices by $A,B,C,\cdots$. 

The supercovariant derivative ${\nabla}$ 
which acts on Lorentz indices is defined as 
\begin{eqnarray}
  {\nabla}_M
  =
  \partial_M-\Omega_M{}^{ab}{\cal O}_{ab}, 
\end{eqnarray}
where 
$\partial_M
=\left(
  \frac{\partial}{\partial x^m}, 
  \frac{\partial}{\partial\theta^\mu}
\right)$ is the left-derivative and ${\cal O}_{ab}$ is the Lorentz 
generator. 
Note that ${\cal O}_{ab}$ has only Lorentz-vector indices. 
The action of $\nabla$ on an Einstein index is defined as follows: 
\begin{eqnarray}
  \nabla_M V_N
  &=&
  (-)^{m(b+n)}E_N{}^B {\nabla}_M V_B
  \nonumber\\
  &=&
  \partial_M V_N + \Gamma_{MN}{}^L V_L. 
\end{eqnarray}
Here, $m,n$ and $b$ are $0$ if $M,N$ and $B$ are bosonic indices 
and $1$ if they are fermionic indices.
The quantity $\Gamma_{MN}{}^L$ is the Christoffel symbol, 
which is chosen to satisfy  
\begin{eqnarray}
  \nabla_M E_N{}^A 
  =
  0. 
\end{eqnarray}
The commutation relation of 
${\nabla}_A=E_A{}^M{\nabla}_M$ is given by 
\begin{eqnarray}
  [{\nabla}_A,{\nabla}_B\}
  =
  R_{AB}{}^{cd}{\cal O}_{cd}
  +
  T_{AB}{}^C{\nabla}_C,  
\end{eqnarray}
where $R_{AB}{}^{cd}$ is the super Riemann tensor and 
$T_{AB}{}^C$ is the supertorsion. 

In \S \ref{subsec:Covariant Derivative as a Set of  Endomorphisms}, 
it is shown that the covariant derivative on a curved space 
can be expressed by a set of matrices. 
In the same way, 
the supercovariant derivative on a superspace 
can be expressed by a set of supermatrices.
 
Consider the principal $G=Spin(d)$ bundle ${\cal E}_{prin}$ 
over the {\it superspace} 
${\cal M}$. 
If ${\cal M}=\cup_i{\cal U}_i$ is an open covering and 
$ z_{[i]}=(x_{[i]},\theta_{[i]})$ and 
$z_{[j]}=(x_{[j]},\theta_{[j]})$ are the coordinates of the same point in 
${\cal U}_i\cap {\cal U}_j$, then 
${\cal E}_{prin}$ is constructed from ${\cal U}_i\times Spin(d)$ 
by identifying $(z_{[i]},g_{[i]})$ and $(z_{[j]},g_{[j]})$ in the case 
$g_{[i]}=t_{ij}(z)g_{[j]}$, where $t_{ij}(z)$ is the transition
function. 
Taking $V=C^\infty({\cal E}_{prin})
=\{f:{\cal E}_{prin}\to{\mathbb C}\}$,   
we can convert the Lorentz 
indices $a$ and $\alpha$ of the supercovariant derivatives into 
those with parentheses:\footnote{
  $R^{({\rm spinor})}{}_{(\alpha)}{}^\beta$ is the appropriate 
  spinor representation. It is Majorana-Weyl for $d=10$ and 
  Majorana for $d=4$. 
}
\begin{eqnarray}
  {\nabla}_{(a)}
  =
  R_{(a)}{}^b(g^{-1}){\nabla}_b, 
  \quad
  {\nabla}_{(\alpha)}
  =
  R_{(\alpha)}{}^\beta(g^{-1}){\nabla}_\beta. 
\end{eqnarray}
As in \S 
\ref{subsec:Covariant Derivative as a Set of Endomorphisms}, 
the derivatives ${\nabla}_{(a)}$ and ${\nabla}_{(\alpha)}$   
are globally defined on ${\cal E}_{prin}$ 
and, indeed, are endomorphisms on $C^\infty({\cal E}_{prin})$. 
The only difference between the two cases 
is that in the present case, transition functions, 
parameters of diffeomorphisms and local Lorentz transformations 
depend not only on the bosonic coordinates $x^m$  
but also on the fermionic coordinates $\theta^\mu$. 
More specificality, ${\nabla}_{(a)}$ 
acts on $f\in V=C^\infty({\cal E}_{prin})$ as follows: 
\begin{eqnarray}
  {\nabla}_{(a)}^{[i]}f^{[i]}(z_{[i]},g_{[i]})
  &=&
  R_{(a)}{}^b(g_{[i]}^{-1})
  {\nabla}_{b}^{[i]}f^{[i]}(z_{[i]},g_{[i]})
  \nonumber\\
  &=&
  R_{(a)}{}^b(g_{[i]}^{-1})
  R_{b}{}^c(t_{ij}(z))
  {\nabla}_{c}^{[j]}f^{[j]}(z_{[j]},g_{[j]})
  \nonumber\\
  &=&  
  R_{(a)}{}^c((t_{ij}(z)^{-1}g_{[i]})^{-1})
  {\nabla}_{c}^{[j]}f^{[j]}(z_{[j]},g_{[j]})
  \nonumber\\
  &=& 
  R_{(a)}{}^c(g_{[j]}^{-1})
  {\nabla}_{c}^{[j]}f^{[j]}(z_{[j]},g_{[j]})
  \nonumber\\
  &=& 
  {\nabla}_{(a)}^{[j]}f^{[j]}(z_{[j]},g_{[j]}). 
\end{eqnarray}
Similarly, we also have
\begin{eqnarray}
  \nabla_{(\alpha)}^{[i]}f^{[i]}(z_{[i]},g_{[i]})
  =
  \nabla_{(\alpha)}^{[j]}f^{[j]}(z_{[j]},g_{[j]}).
\end{eqnarray}
Therefore, the ${\nabla}_{(a)}$ and ${\nabla}_{(\alpha)}$
are endomorphisms on a supervector space 
$C^\infty({\cal E}_{prin})$, which are described by supermatrices, 
as we explain in the next section. 

Before concluding this subsection, 
we give a proof that ${\nabla}_{(a)}$ 
and ${\nabla}_{(\alpha)}$ are anti-Hermitian and Hermitian, 
respectively. The standard inner product is defined by 
\begin{eqnarray}
  \left(u(z,g),v(z,g)\right)
  =
  \int dg\int E\hspace{0.05cm}
  d^dx\hspace{0.05cm} d^{d_s}\theta
  \hspace{0.05cm}u^\ast(z,g)v(z,g), 
\end{eqnarray}
where $dg$ is the Haar measure of $Spin(d)$ and 
$E=sdet E_M{}^A$. 
Then, we can show the (anti-) Hermiticity of $\nabla_{(A)}$ in the following way: 
\begin{eqnarray}
  \left(u_{(a)}(z,g),{\nabla}^{(a)}v(z,g)\right)
  &=&
  \int E\hspace{0.05cm}
  d^dx\hspace{0.05cm} 
  d^{d_s}\theta\hspace{0.05cm}
  u_{(a)}^\ast(z,g){\nabla}^{(a)}v(z,g)
  \nonumber\\
  &=&
  \int E\hspace{0.05cm}
  d^dx\hspace{0.05cm}
  d^{d_s}\theta\hspace{0.05cm}
  u_{a}^\ast(z,g){\nabla}^{a}v(z,g)
  \nonumber\\
  &=&
  -\int E\hspace{0.05cm}
  d^dx\hspace{0.05cm}
  d^{d_s}\theta\hspace{0.05cm}
  \left({\nabla}^{a}u_{a}^\ast(z,g)\right)v(z,g)
  \nonumber\\
  &=&
  -\int E\hspace{0.05cm}
  d^dx\hspace{0.05cm}
  d^{d_s}\theta\hspace{0.05cm}
  \left({\nabla}^{(a)}u_{(a)}^\ast(z,g)\right)v(z,g)
  \nonumber\\
  &=&
  -\left({\nabla}^{(a)}u_{(a)}(z,g),v(z,g)\right). 
\end{eqnarray}
Similarly, we can show that 
\begin{eqnarray}
  \left(u_{(\alpha)}(z,g),{\nabla}^{(\alpha)}v(z,g)\right)
  =
  \left({\nabla}^{(\alpha)}u_{(\alpha)}(z,g),v(z,g)\right).
\end{eqnarray}
The extra minus sign appears because $u^\alpha$ is fermionic. 

The method presented in this subsection can be applied to 
a supersymmetry with any value of ${\cal N}$. 
\subsection{Example:two-dimensional ${\cal N}=(1,1)$ superspace}
As a concrete example, let us construct a two-dimensional 
${\cal N}=(1,1)$ superspace ${\cal M}$ with Lorentzian signature and 
a principal $Spin(1,1)$ bundle ${\cal E}_{prin}$ over it.  

In order to eliminate the fields that are unnecessary in 
supergravity, 
we impose the following torsion constraints \cite{Howe}: 
\begin{eqnarray}
  T_{ab}{}^c=T_{\alpha\beta}{}^\gamma=0, 
  \qquad
  T_{\alpha\beta}{}^c
  =2i(C\gamma^c)_{\alpha\beta}. 
\end{eqnarray}
Here $C_{\alpha\beta}$ is the charge-conjugation matrix. 
Let $M$ be the ordinary manifold which is obtained by setting 
all the fermionic coordinates  $\theta$ of $M$ to zero.  
We denote by $e_a{}^m(x)$ and $\omega_m(x)=\omega_m{}^{01}(x)$ 
the zweibein on $M$ and the torsionless spin connection 
associated with $e_a{}^m(x)$, respectively. 
Then, using the higher-$\theta$ components of the
superdiffeomorphism symmetry, we can write the superzweibein $E(x,\theta)$ 
and the spin connection $\Omega(x,\theta)$ in the form of 
the Wess-Zumino gauge \cite{Howe}: 
\begin{eqnarray}
  E_M{}^A(x,\theta)
  &=&
  \left(
    \begin{array}{cc}
      e_m{}^a(x) & -\frac{1}{2}\omega_m(x)(\gamma_{01}\theta)^\alpha\\
      i(C\gamma^a\theta)_\mu
      & \delta_\mu{}^\alpha
    \end{array}
  \right), 
  \nonumber\\
  E_A{}^M(x,\theta)
  &=&
  \left(
    \begin{array}{cc}
      e_a{}^m(x) & \frac{1}{2}\omega_a(x)(\gamma_{01}\theta)^\mu\\
      -ie_b{}^m(x)(C\gamma^b\theta)_\alpha
      & \delta_\alpha{}^\mu
      -\frac{i}{4}(\bar{\theta}\theta)
      (\gamma_{01}\gamma^b)^\mu{}_\alpha\omega_b(x)
    \end{array}
  \right), 
\end{eqnarray}
\begin{eqnarray}
  \Omega_{m}{}^{01}(x,\theta)
  =
  \Omega_m(x,\theta)
  =
  \omega_m(x), 
  \qquad
  \Omega_{\mu}^{01}(x,\theta)
  =
  \Omega_\mu(x,\theta)=0. 
\end{eqnarray}
Here we have set the gravitino and auxiliary fields to zero for simplicity. 
The supercovariant derivative ${\nabla}_A=E_A{}^M{\nabla}_M$ is given
by 
\begin{eqnarray}
  & &
  {\nabla}_a
  =
  E_a{}^M{\nabla}_M
  =
  e_a{}^m\partial_m
  -
  2\omega_a
  \left\{
    {\cal O}_{01}
    -
    \frac{1}{4}(\gamma_{01}\theta)^\mu\partial_\mu 
  \right\}, 
  \label{covariant derivative on superspace:vector}
  \\
  & &
  {\nabla}_\alpha
  =
  E_\alpha{}^M{\nabla}_M
  =
  \delta_\alpha{}^\mu\partial_\mu
  -
  i(C\gamma^b\theta)_\alpha{\nabla}_b. 
  \label{covariant derivative on superspace:spinor}
\end{eqnarray}
Although ${\cal O}_{01}$ does not act on the index $\mu$, 
if we regard ${\cal O}_{01}
-\frac{1}{4}(\gamma_{01}\theta)^\mu\partial_\mu $ as 
a ``Lorentz generator'', it does act on $\mu$, and therefore $\theta^\mu$ 
can be regarded as a ``Lorentz spinor''. 
Then ${\nabla}_a$ can be regarded as the ordinary covariant derivative on $M$ 
which also acts on the index $\mu$. 
In this sense, ${\cal M}$ is a spinor bundle over $M$ 
with Grassmann-odd fiber coordinates $\theta^\mu$, 
which is associated with the spin structure. 

Now we introduce a principal $Spin(1,1)$ bundle ${\cal E}_{prin}$ 
over ${\cal M}$ associated with the spin structure. Then, 
we regard the supercovariant derivative ${\nabla}_A$ 
as acting on ${\cal E}_{prin}$.  
Parameterizing the spin $s$ representation of $Spin(1,1)$ as  
\begin{eqnarray}
  R^{\langle s\rangle}(\varphi)
  =
  e^{2s\varphi},  
\end{eqnarray}
we can express the Lorentz generator as 
\begin{eqnarray}
  {\cal O}_{01}
  =
  \frac{1}{4}\partial_{\varphi}.  
\end{eqnarray}
Then, if we introduce $\theta^{(\mu)}$ according to the relation  
\begin{eqnarray}
  \theta^{(\mu)}
  =
  (e^{\varphi\gamma_{01}})^{(\mu)}{}_\nu\theta^{\nu},   
  \label{change of the coordinate of Eprin}
\end{eqnarray}
$\theta^{(\mu)}$ is globally defined on ${\cal E}_{prin}$. 
Next, using $x^m,\theta^{(\mu)}$ and $\varphi$ as independent variables, 
we can rewrite Eqs. (\ref{covariant derivative on superspace:vector}) 
and (\ref{covariant derivative on superspace:spinor}) as  
\begin{eqnarray}
  {\nabla}_a
  &=&
  e_a{}^{m}\left(
    \partial_{m}
    -
    \frac{1}{2}\omega_{m}\partial_{\varphi}
  \right), 
  \nonumber\\
  {\nabla}_\alpha
  &=&
  (e^{\varphi\gamma_{01}})^{(\mu)}{}_\alpha
  \partial_{(\mu)}
  -
  i(C\gamma^b
  e^{\varphi\gamma_{01}}
  \theta)_\alpha{\nabla}_b.  
\end{eqnarray}
Finally, as explained in the previous subsection,
we can convert indices without parentheses into those with
parentheses, and hence we have 
\begin{eqnarray}
  {\nabla}_{(a)}
  &=&
  e_{(a)}{}^{m}
  \left(
    \partial_{m}
    -
    \frac{1}{2}\omega_{m}
    \partial_{\varphi}
  \right), 
  \label{parenthesized cov.deriv.vector}
  \\
  {\nabla}_{(\alpha)}
  &=&
  \delta_{(\alpha)}{}^{(\mu)}
  \partial_{(\mu)}
  -
  i\left(C\gamma^{(b)}\theta\right)_{(\alpha)}{\nabla}_{(b)}. 
  \label{parenthesized cov.deriv.spinor}
\end{eqnarray}
Note that ${\cal E}_{prin}$ can be constructed 
as a direct product of 
${\mathbb R}^{0|2}=\{\theta^{(1)},\theta^{(2)}\}$ 
and $E_{prin}$. 
\\
\\
\noindent
\textbf{Supersphere}\\

We now construct the homogeneous and isotropic Euclidean 
supersphere $S^{2|2}$ and the covariant derivative on it. 
Let us start with the case of the ordinary $S^2$ \cite{HKK}.  
Then, by parameterizing $Spin(2)$ as  
\begin{eqnarray}
  R^{\langle s\rangle}(\varphi)
  =
  e^{2is\varphi}, 
  \quad
  (\varphi\in [0,2\pi))  
\end{eqnarray}
we can express the Lorentz generator in terms of 
the derivative with respect to $\varphi$ as
\begin{eqnarray}
  {\cal O}_{+-}
  =
  \frac{1}{4i}\partial_\varphi, 
\end{eqnarray}
where $+$ and $-$ indicate the linear combinations of the 
Lorentz indices $1+i2$ and $1-i2$, respectively. 

In the stereographic coordinates $(w,\bar{w})$ 
projected from the north pole, the metric is 
\begin{eqnarray}
  g_{w\bar{w}}=g_{\bar{w}w}=
  \frac{1}{(1+w\bar{w})^2}, 
  \qquad
  g_{ww}=g_{\bar{w}\bar{w}}=0, 
\end{eqnarray}
and we can take the zweibein and spin connection as 
\begin{eqnarray}
  e_{+\bar{w}}
  =
  e_{-w}
  =
  \frac{1}{1+w\bar{w}}, 
  \quad
  e_{+w}
  =
  e_{-\bar{w}}
  =  
  0, 
  \\
  \omega_w{}^{+-}
  =
  \frac{\bar{w}}{1+w\bar{w}}, 
  \quad
  \omega_{\bar{w}}{}^{+-}
  =
  -\frac{w}{1+w\bar{w}}.  
\end{eqnarray}
Therefore, the covariant derivatives with parentheses 
are given by
\begin{eqnarray}
  \tilde{{\nabla}}_{(+)}^{[w]}
  &=&
  e^{-2i\varphi}
  \left(
    (1+w\bar{w})\partial_w
    +
    \frac{i}{2}\bar{w}\partial_\varphi 
  \right), 
  \nonumber\\
  \tilde{{\nabla}}_{(-)}^{[w]}
  &=&
  e^{2i\varphi}
  \left(
    (1+w\bar{w})\partial_{\bar{w}}
    -
    \frac{i}{2}w\partial_\varphi 
  \right).   
 \label{Cov.deriv_north}
\end{eqnarray}
In the same way, in the stereographic coordinates $(w^\prime,\bar{w}^\prime)$ 
projected from the south pole, which are related to $(w,\bar{w})$ 
as $w^\prime=1/w$, we have 
\begin{eqnarray}
  \tilde{{\nabla}}_{(+)}^{[w^\prime]}
  &=&
  e^{-2i\varphi^\prime}
  \left(
    (1+w^\prime\bar{w}^\prime)\partial_{w^\prime}
    +
    \frac{i}{2}\bar{w}^\prime\partial_{\varphi^\prime} 
  \right), 
  \nonumber\\
  \tilde{{\nabla}}_{(-)}^{[w^\prime]}
  &=&
  e^{2i\varphi^\prime}
  \left(
    (1+w^\prime\bar{w}^\prime)\partial_{\bar{w}^\prime}
    -
    \frac{i}{2}w^\prime\partial_{\varphi^\prime} 
  \right).   
  \label{Cov.deriv_south}
\end{eqnarray}
Here, the coordinate $\varphi^\prime$ of the fiber on this patch is 
related to $\varphi$ according to the transition function 
\begin{eqnarray}
  \varphi^\prime
  =
  \varphi+\arg(w)+\frac{\pi}{2}. 
  \label{transition function_sphere}
\end{eqnarray}
We can explicitly check that Eqs. (\ref{Cov.deriv_north}) 
and (\ref{Cov.deriv_south}) are identical. 
Note that $E_{prin}$ on $S^2$ is $Spin(2)(=S^1)$ bundle over 
$S^2$ and is topologically $S^3$. 

Next we consider the supersphere $S^{2|2}$ and  
the principal $Spin(2)$ bundle ${\cal E}_{prin}$ over it.  
In this case Eqs. (\ref{parenthesized cov.deriv.vector}) 
and (\ref{parenthesized cov.deriv.spinor}) become 
\begin{eqnarray}
  {\nabla}_{(+)}^{[w]}
  &=&
  e^{-2i\varphi}
  \left\{
    (1+w\bar{w})\partial_w
    +
    \frac{i}{2}\bar{w}
    \partial_{\varphi}
  \right\}, 
  \nonumber\\
  {\nabla}_{(-)}^{[w]}
  &=&
  e^{2i\varphi}
  \left\{
    (1+w\bar{w})\partial_{\bar{w}}
    -
    \frac{i}{2}w
    \partial_{\varphi}
  \right\}, 
  \\
  {\nabla}_{(\alpha)}^{[w]}
  &=&
  \delta_{(\alpha)}{}^{(\mu)}\partial_{(\mu)}
  -
  i\left(C\gamma^{(b)}\theta\right)_{(\alpha)}{\nabla}_{(b)}. 
\end{eqnarray}
Using the relation 
$(w,\theta_{[w]}^{(\mu)})=
(1/w^\prime,\theta_{[w^\prime]}^{(\mu)})$ 
and Eq. (\ref{transition function_sphere}), 
we can explicitly confirm the relation   
\begin{eqnarray}
  {\nabla}_{(A)}^{[w]}={\nabla}_{(A)}^{[w^\prime]}. 
\end{eqnarray}
In this way, we can describe the supersphere $S^{2|2}$ 
in terms of four supermatrices. 
\section{Supermatrix model and local supersymmetry}
\label{sec:Supermatrix model}
In the previous section, we showed that 
the supercovariant derivative on a superspace can be 
described by a set of supermatrices. 
Therefore, 
as in \S \ref{subsec:New interpretation of IIB matrix model}, 
we can embed local supersymmetry manifestly in 
supermatrix models.  
In this section, we consider a straightforward 
generalization of IIB matrix model. 
We show that the equation of motion of the supermatrix model is 
compatible with that of supergravity. 
\subsection{Supermatrix model}
An element $f:{\cal E}_{prin}\to{\mathbb C}$ 
of $V=C^\infty({\cal E}_{prin})$ can be expressed 
as a power series in $\theta^\mu$. 
The coefficients of even powers of $\theta^\mu$ 
in this series 
are Grassmann-even functions and form the 
Grassmann-even subspace $V_e$ of $V$. 
Similarly, the coefficients of odd powers of $\theta^\mu$ 
form the Grassmann-odd subspace $V_o$.  
Therefore, the total space $V$ is the direct sum of $V_e$ and $V_o$: 
\begin{eqnarray}
  V=V_e\oplus V_o. 
\end{eqnarray}
If we introduce a regularization of $V$ in which 
the dimensions of $V_e$ and $V_o$ are $N_e$ and $N_o$, 
respectively, then a linear transformation on $V$ 
can be expressed as a $(N_e,N_o)$-supermatrix. 
We summarize the definition and 
properties of supermatrix in Appendix \ref{appendix:supermatrix}. 
 
The derivatives ${\nabla}_{(a)}$ and ${\cal O}_{ab}$  
map an even (resp., odd) element to an even (resp., odd) element.  
Hence they can be expressed in terms of even supermatrices. 
Contrastingly, ${\nabla}_{(\alpha)}$ 
maps an even element to an odd element and vice versa. 
Therefore they are expressed in terms of odd supermatrices. 

Now  we  propose the supermatrix generalization of IIB matrix
model. 
The dynamical variables consist of Hermitian even supermatrices 
${\cal A}_{a}$ with vector indices 
and the Hermitian odd supermatrices 
$\Psi_{\alpha}$ with spinor indices.  
The action is given by 
\begin{eqnarray}
  S=-\frac{1}{g^2}Str\left(
    \frac{1}{4}[{\cal A}_{a},{\cal A}_{b}][{\cal A}^{a},{\cal A}^{b}]
    +
    \frac{1}{2}
    \bar{\Psi}^{\alpha}\left(\gamma^{a}\right)_{\alpha}{}^{\beta}
    [{\cal A}_{a},\Psi_{\beta}]
  \right). 
  \label{action:superIIB}
\end{eqnarray}
This model possesses superunitary symmetry $U(N_e|N_o)$. 
If the matrices ${\cal A}_{a}$ are sufficiently close to the 
covariant derivatives $i{\nabla}_{(a)}$ on some ${\cal E}_{prin}$, we regard 
the $A_{a}$ and $\Psi_{\alpha}$ as endomorphisms of $V=C^\infty({\cal E}_{prin})$. 
As we discuss in the next section, 
there is a possibility that Eq. (\ref{action:superIIB}) is equivalent to
the original IIB matrix model.   

A remark on the role of the odd supermatrix $\Psi_\alpha$ is in order
here. The action (\ref{action:superIIB}) has the global symmetry  
\begin{eqnarray}
	\delta^{(1)}{\cal A}_a
	=
	-i\overline{\Psi}\gamma_a\hat{\epsilon}, 
	\qquad
	\delta^{(1)}\Psi_\alpha
	=
	\frac{i}{2}
	\left[{\cal A}_a,{\cal A}_b\right]
	\left(\gamma^{ab}\hat{\epsilon}\right)_\alpha
        \label{global SUSY 1}
\end{eqnarray}
and 
\begin{eqnarray}
	\delta^{(2)}{\cal A}_a
	=
	0, 
	\qquad
	\delta^{(2)}\Psi_\alpha
	=
	\hat{\epsilon}_\alpha, 
        \label{global SUSY 2}
\end{eqnarray}
which is the analogue of the ${\cal N}=2$ global supersymmetry 
of the original IIB matrix model. 
Here, the quantities 
\\
\begin{eqnarray}
 & &
  \hspace{1.7cm}
  \overset{N_e}{\rotatebox{90}{\}}}
  \hspace{0.35cm}
  \overset{N_o}{\rotatebox{90}{\}}}
  \nonumber\\
   & &
  \hat{\epsilon}_\alpha
  =
  \epsilon_\alpha\left(
    \begin{array}{cc}
      1 & 0\\
      0 & -1
    \end{array}
  \right)
  \begin{array}{c}
    {\scriptstyle \}N_e }\\
    {\scriptstyle \}N_o }
   \end{array}
\end{eqnarray}
are odd {\it scalar} supermatrices, 
which commute with ${\cal A}_a$ and anticommute with $\Psi_\alpha$.    
In the original interpretation of 
IIB matrix model, this symmetry is simply the ten-dimensional 
${\cal N}=2$ supersymmetry. 
In the case of the supermatrix model, however,  
the local supersymmetry is manifestly embedded 
as a part of the superunitary symmetry,  
as we see in \S \ref{subsec:Local Supersymmetry}.  
Indeed, a supermatrix model without $\Psi_\alpha$, such as 
\begin{equation}
   S=-\frac{1}{4g^2}Str
   \left(
     [{\cal A}_a,{\cal A}_b][{\cal A}^a,{\cal A}^b]
   \right), 
   \label{action:superIIB_bosonic}
\end{equation}
also possesses this local supersymmetry. 
Furthermore, as we see in the next subsection, 
both Eqs. (\ref{action:superIIB}) and (\ref{action:superIIB_bosonic}) 
describe supergravity at the level of classical solutions.  
On the other hand, in the original interpretation of IIB matrix model, 
in which matrices are regarded as coordinates, 
the ${\cal N}=2$ global supersymmetry prevents the eigenvalues 
from collapsing to one point, and this allows a reasonable interpretation 
of spacetime \cite{IKKT}. 
We conjecture that the global symmetry 
represented by Eqs. (\ref{global SUSY 1}) and 
(\ref{global SUSY 2}) plays a similar role here 
and allows ${\cal A}_a$ to fluctuate around $i\nabla_{(a)}$ 
on some manifold.   
\subsection{Classical solutions}
\label{subsec:Classical Solutions_super}
In this subsection we investigate classical solutions of 
Eq. (\ref{action:superIIB}). 
For simplicity, let us consider the four-dimensional 
model here. We use the notation of Ref. \citen{WZ77}. 
In this case,  
${\cal A}_{a}$ and $\Psi_{\alpha}$ are four-dimensional
vector and Majorana spinor, respectively.  
The equations of motion are given by 
\begin{eqnarray}
  & &
  [{\cal A}^a,[{\cal A}_a,{\cal A}_b]]
  +
  (C\gamma_b)^{\alpha\beta}\{\Psi_\alpha,\Psi_\beta\}
  =
  0, \label{eq:EOM1}
  \\
  & &
  (\gamma^a)_\alpha{}^\beta[{\cal A}_a,\Psi_\beta]
  =
  0, \label{eq:EOM2}
\end{eqnarray}
where $C^{\alpha\beta}$ is the charge-conjugation matrix. 

We impose the following ansatz:\footnote{
If we start with ${\cal A}_a=i{\nabla}_{(a)}$ 
and $\Psi_\alpha=\kappa{\nabla}_{(\alpha)}$, where $\kappa$ is a constant, 
then we can show that $\kappa$ must be zero. 
(See Appendix \ref{subsec:kappa=0}.) 
}
\begin{eqnarray}
  {\cal A}_a=i{\nabla}_{(a)}, 
  \qquad
  \Psi_\alpha=0. 
  \label{ansatz:matrix and covariant derivative}
\end{eqnarray}
In the bosonic case, once we impose the ansatz 
(\ref{ansatz:bosonic case}), we can obtain the general solution of the
matrix equation of motion (\ref{eq:EOM with parenthesis_1}). 
In the present case, however, it is not easy to find the general solution of 
Eqs. (\ref{eq:EOM1}) and (\ref{eq:EOM2}). 
Instead, we show that the ansatz (\ref{ansatz:matrix and covariant derivative}) 
satisfies Eqs. (\ref{eq:EOM1}) and (\ref{eq:EOM2})  
if we impose the standard off-shell torsion constraints \cite{GWZ78,WB}
\begin{eqnarray}
  T_{\alpha\beta}{}^c
  &=&
  2i(\gamma^c C^{-1})_{\alpha\beta},
  \nonumber\\
  T_{\alpha\beta}{}^\gamma
  &=&
  0, 
  \nonumber\\
  T_{a\beta}{}^c
  &=&
  0, 
  \nonumber\\
  T_{ab}{}^c
  &=&
  0, 
  \label{eq:torsion+EOM1}
\end{eqnarray}
and the equation of motion of supergravity \cite{GWZ78}, 
\begin{eqnarray}
  T_{a\beta}{}^\gamma 
  &=&
  0. 
  \label{eq:torsion+EOM5}
\end{eqnarray} 
Using Eqs. (\ref{eq:torsion+EOM1}) and (\ref{eq:torsion+EOM5}), we have 
\begin{eqnarray}
  [{\nabla}^a,[{\nabla}_a,{\nabla}_b]]
  &=&
  [{\nabla}^a,R_{ab}{}^{cd}{\cal O}_{cd}+T_{ab}{}^\gamma{\nabla}_\gamma]
  \nonumber\\
  &=&
  ({\nabla}^a R_{ab}{}^{cd}){\cal O}_{cd}
  -
  R_{ab}{}^{ac}{\nabla}_c
  +
  ({\nabla}^a T_{ab}{}^\gamma){\nabla}_\gamma
  +
  T_{ab}{}^\gamma R^a{}_{\gamma}{}^{cd}{\cal O}_{cd}.  
  \label{eq:EOM3}
\end{eqnarray}
Furthermore, as shown in Appendix \ref{sec:Derivation of EOM}, 
from Eqs. (\ref{eq:torsion+EOM1}), (\ref{eq:torsion+EOM5}) and 
the Bianchi identities, we can show the relations   
\begin{eqnarray}
  R_{ab}{}^{ac}
  &=&
  0, 
  \label{eq:EOM_component_1}
  \\
  {\nabla}^a T_{ab}{}^\gamma
  &=&
  0, 
  \label{eq:EOM_component_2}
  \\
  {\nabla}^a R_{ab}{}^{cd}
  &=&
  0, 
  \label{eq:EOM_component_3}
  \\  
  T_{ab}{}^\gamma R^a{}_{\gamma}{}^{cd}
  &=&
  0,    
  \label{eq:EOM_component_4}
\end{eqnarray}
which indicate that the right-hand side of Eq. (\ref{eq:EOM3}) is zero. 
Hence, the ansatz 
(\ref{ansatz:matrix and covariant derivative}) 
with the constraints 
(\ref{eq:torsion+EOM1}) and (\ref{eq:torsion+EOM5}) satisfies 
Eqs. (\ref{eq:EOM1}) and (\ref{eq:EOM2}). 

Some comments are in order here. 
First, although we showed that Eqs. (\ref{eq:torsion+EOM1}) and 
(\ref{eq:torsion+EOM5}) are sufficient conditions 
for Eqs. (\ref{eq:EOM1}) and (\ref{eq:EOM2}) 
under the assumption  (\ref{ansatz:matrix and covariant derivative}),  
it is unclear whether or not they are necessary.  
Secondly, it is desirable to determine whether 
the torsion constraints in the ten-dimensional superspace  
\cite{ADR} are compatible with the equation of motion of the supermatrix model. 
Thirdly, because we have set $\Psi_\alpha$ 
to zero as an ansatz, 
the action (\ref{action:superIIB_bosonic}) allows 
the same classical solutions. 
\subsection{Superdiffeomorphism and local supersymmetry}
\label{subsec:Local Supersymmetry}
The supermatrix model possesses the superunitary symmetry
\begin{eqnarray}
  \delta{\cal A}_a
  =
  i[\Lambda,{\cal A}_a], 
  \quad
  \delta\Psi_\alpha
  =
  i[\Lambda,\Psi_\alpha],  
  \label{super unitary transformation}
\end{eqnarray}
where $\Lambda$ is an even Hermitian supermatrix. 
If we expand ${\cal A}_a$ and $\Psi_\alpha$ as\footnote{
As in \S \ref{subsubsec:Diffeo and LL},  
because ${\cal A}_a$ and $\Psi_\alpha$ are Hermitian, 
we should introduce the commutator and the anticommutator.  
Here we omit them for simplicity. } 
\begin{eqnarray}
  {\cal A}_a
  &=&
  i{\nabla}_{(a)}
  +
  a_{(a)}(z,g)
  +
  ia_{(a)}{}^{(B)}(z,g){\nabla}_{(B)}
  +
  ia_{(a)}{}^{bc}(z,g){\cal O}_{bc}
  \nonumber\\
  & &
  \quad
  +
  i^2a_{(a)}{}^{(B)(C)}(z,g){\nabla}_{(B)}{\nabla}_{(C)}
  +
  i^2a_{(a)}{}^{(B),cd}(z,g){\nabla}_{(B)}{\cal O}_{cd}
  +
  \cdots, 
  \nonumber\\
  \Psi_\alpha
  &=&
  \psi_{(\alpha)}(z,g)
  +
  i\psi_{(\alpha)}{}^{(B)}(z,g){\nabla}_{(B)}
  +
  i\psi_{(\alpha)}{}^{bc}(z,g){\cal O}_{bc}
  \nonumber\\
  & &
  \quad
  +
  i^2\psi_{(\alpha)}{}^{(B)(C)}(z,g){\nabla}_{(B)}{\nabla}_{(C)}
  +
  i^2\psi_{(\alpha)}{}^{(B),cd}(z,g){\nabla}_{(B)}{\cal O}_{cd}
  +
  \cdots, 
  \nonumber\\
  {\nabla}_{(A)}
  &=&
  E_{(A)}{}^M{\nabla}_M, 
  \label{expansion by local fields:super version1}
\end{eqnarray}
and take $\Lambda$ as 
\begin{eqnarray}
  \Lambda
  =
  \frac{i}{2}
  \left\{
    \lambda^{(a)}(z),{\nabla}_{(a)}
  \right\}
  +
  \frac{i}{2}
  \left[
    \lambda^{(\alpha)}(z),{\nabla}_{(\alpha)}
  \right], 
\end{eqnarray}
then Eq. (\ref{super unitary transformation}) becomes 
a superdiffeomorphism 
on the fields that appear in 
Eq. (\ref{expansion by local fields:super version1}).  

The part of the superdiffeomorphism generated by 
\begin{eqnarray}
   \Lambda
  =
  \frac{i}{2}
  \left[
    \lambda^{(\alpha)}(x),{\nabla}_{(\alpha)}
  \right]
  \label{generator of local SUSY}
\end{eqnarray}
gives the local supersymmetry. 
For example, in the Wess-Zumino gauge \cite{WZ77,WB} 
\begin{eqnarray}
  \left.
    E_M{}^A
  \right|_{\theta=0}
  =
  \left(
    \begin{array}{cc}
      e_m{}^a(x) & \frac{1}{2}\chi_m{}^\alpha(x)\\
      0 & \delta_\mu{}^\alpha
    \end{array}
  \right), 
\end{eqnarray}
where $e_m{}^a$ is the vielbein 
and $\chi_m{}^\alpha$ is the gravitino, 
Eq. (\ref{generator of local SUSY}) gives  
\begin{eqnarray}
  \delta e_m{}^a(x)
  =
  2i\bar{\lambda}(x)\gamma^a\chi_\mu(x), 
  \qquad
  \delta\chi_m{}^\alpha(x)
  =
  2{\nabla}_m\lambda^\alpha(x)
  +
  ({\rm auxiliary\ fields}). 
\end{eqnarray}
\section{Conclusions and discussion}
\label{sec:Conclusions and Discussions}
In Ref. \citen{HKK} we showed that the covariant derivative 
on a $d$-dimensional curved space can be described 
by a set of $d$ matrices. 
Based on this, we introduced a new interpretation of the matrix model, 
in which we regard matrices as {\it covariant derivatives} 
on a curved space rather than {\it coordinates}. 
With this interpretation, 
the symmetries of general relativity are 
included in the unitary symmetry of the matrix model, and 
the path integral contains a summation over all curved spaces.  
In this paper, we have extended this formalism to 
include supergravity. 
If we promote matrices to {\it supermatrices}, we can express  
{\it supercovariant derivatives}  
as special configurations of the supermatrices. 
Then, the local supersymmetry is included in the superunitary symmetry. 
Furthermore, we showed that if we consider the action  
$S=-Str\left[{\cal A}_a,{\cal A}_b\right]^2+\cdots$,  
the matrix equation of motion follows from 
the Einstein and the Rarita-Schwinger equations  
in the four-dimensional case. 

There remain several problems. 
First, it is desirable to clarify the relationship between   
our new interpretation and the original one, 
in which matrices represent coordinates. 
Although gravity is realized manifestly 
in the new interpretation, 
its relationship to string theory is rather obscure. 
Contrastingly, 
the original interpretation is directly connected to string theory, 
because the action can be regarded as 
the Green-Schwarz action of a type-IIB superstring. 
Furthermore, we have not understood the meaning of 
the global ${\cal N}=2$ supersymmetry of IIB matrix model  
in the new interpretation. 
Because we have realized supersymmetry as 
superunitary symmetry, it seems that 
global ${\cal N}=2$ supersymmetry need not exist. 
In the original IIB matrix model, however,  
it prevents the eigenvalues from collapsing to a point 
and allows a reasonable interpretation of spacetime \cite{IKKT}. 
We believe that in the new interpretation  
it plays a similar role and allows 
${\cal A}_a$ to fluctuate around a covariant derivative on some
manifold. 
In order to connect the two different interpretations,  
it may be helpful to consider noncommutative geometry. 
This is because there is no essential difference between 
coordinates and derivatives in noncommutative geometry 
\cite{hep-th/9612222}. 
If we can understand the relationship more closely, 
the role of the global ${\cal N}=2$ supersymmetry 
may become clear. 
Furthermore, this would be useful for 
the construction of curved noncommutative manifolds. 

Secondly, it remains to 
investigate whether ten-dimensional supergravity 
can be derived from 
the equation of motion of our supermatrix model. 
In Refs. \citen{ADR}, the superspace formulation of 
ten-dimensional ${\cal N}=1$ supergravity coupled to 
super Yang-Mills was studied, and the torsion constraints 
were presented. 
The calculation in the ten-dimensional case is more 
complicated,  
due to the presence of various fields, including 
the dilaton, antisymmetric tensor and gauge field. 
At present, it is unclear 
whether the torsion constraints presented in Refs. \citen{ADR} 
fit our supermatrix model. 
If they do not, we may need to modify the form of the action or 
to find other torsion constraints. 

Thirdly, the supermatrix model we have proposed might not be 
well-defined, because the action is not bounded from below, 
due to the minus sign appearing in the supertrace. 
The possibility that supermatrix models 
are formally equivalent to the corresponding 
ordinary matrix models is discussed in Refs. \citen{Alvarez-Gaume} 
and \citen{OT}. 
\footnote{
As the action of the original IIB matrix model 
is formally the same as the effective 
action of D-instantons, 
our supermatrix model (\ref{action:superIIB}) 
is identical to the effective action of D-instantons and 
ghost D-instantons \cite{OT}.   
}  
If a similar mechanism holds in our supermatrix model, 
we may be able to embed supergravity  
in the original IIB matrix model. 
\section*{Acknowledgements}
The authors would like to thank M. Bagnoud, F. Kubo, 
Y. Matsuo, A. Miwa, K. Murakami and Y. Shibusa 
for stimulating discussions and comments.   
The work of M. H. and Y. K. was supported in part 
by JSPS Research Fellowships for Young Scientists. 
This work was also supported in part by a Grant-in-Aid for
the 21st Century COE ``Center for Diversity and Universality in
Physics''. 
\renewcommand{\theequation}{\Alph{section}.\arabic{equation}}
\appendix
\section{Supermatrices}
\label{appendix:supermatrix}
\setcounter{equation}{0} 
In this appendix we define a supermatrix and summarize its properties. 

Consider a supervector space $V$. 
We use the standard basis in which 
supervectors have the form 
\begin{eqnarray}
  v=\left(
    \begin{array}{c}
      v_1\\
      v_2
    \end{array}
  \right)
  \begin{array}{c}
    {\scriptstyle \}N_e }\\
    {\scriptstyle \}N_o }
  \end{array}, 
\end{eqnarray}
where $v_1$ and $v_2$ are bosonic and fermionic column vectors, respectively. 
We call $v$ an {\it even} supervector. 
Multiplying $v$ by a Grassmann-odd number, we obtain   
an {\it odd} supervector. 
  
In the standard basis defined above, 
an {\it even} supermatrix ${\cal B}$ 
and an {\it odd} supermatrix ${\cal F}$ can be written as 
\begin{eqnarray}
  & &
  \hspace{1.3cm}
  \overset{N_e}{\rotatebox{90}{\}}}
  \hspace{0.5cm}
  \overset{N_o}{\rotatebox{90}{\}}}
  \hspace{3.85cm}
  \overset{N_e}{\rotatebox{90}{\}}}
  \hspace{0.5cm}
  \overset{N_o}{\rotatebox{90}{\}}}
  \nonumber\\
   & &
  {\cal B}
  =
  \left(
    \begin{array}{cc}
      B_1 & F_1\\
      F_2 & B_2
    \end{array}
  \right)
  \begin{array}{c}
    {\scriptstyle \}N_e }\\
    {\scriptstyle \}N_o }
  \end{array}, 
  \qquad
  {\cal F}
  =
  \left(
    \begin{array}{cc}
      F_1^\prime & B_1^\prime\\
      B_2^\prime & F_2^\prime
    \end{array}
  \right)
  \begin{array}{c}
    {\scriptstyle \}N_e }\\
    {\scriptstyle \}N_o }
    \end{array}. 
\end{eqnarray}
Here $B$ and $B^\prime$ are bosonic matrices and $F$ and $F^\prime$ are 
fermionic matrices. 
An even supermatrix represents a linear transformation 
of the supervectorspace $V$, which maps an even (resp., odd) supervector 
to an even (resp., odd) supervector. 
An odd supermatrix also represents a linear transformation 
of $V$, but it maps 
an even (resp., odd) supervector to an odd (resp., even) supervector.  

A {\it scalar supermatrix} is defined by 
\begin{eqnarray}
 & &
  \hspace{1.35cm}
  \overset{N_e}{\rotatebox{90}{\}}}
  \hspace{0.19cm}
  \overset{N_o}{\rotatebox{90}{\}}}
  \hspace{3.7cm}
  \overset{N_e}{\rotatebox{90}{\}}}
  \hspace{0.3cm}
  \overset{N_o}{\rotatebox{90}{\}}}
  \nonumber\\
   & &
  \hat{c}
  =
  c\left(
    \begin{array}{cc}
      1 & 0\\
      0 & 1
    \end{array}
  \right)
  \begin{array}{c}
    {\scriptstyle \}N_e }\\
    {\scriptstyle \}N_o }
  \end{array}, 
  \qquad
  \hat{\epsilon}
  =
  \epsilon\left(
    \begin{array}{cc}
      1 & 0\\
      0 & -1
    \end{array}
  \right)
  \begin{array}{c}
    {\scriptstyle \}N_e }\\
    {\scriptstyle \}N_o }
   \end{array}, 
\end{eqnarray}
where $c$ and $\epsilon$ are Grassmann-even and odd numbers. 
Then, 
$\hat{c}$ commutes with both even and odd supermatrices, and 
$\hat{\epsilon}$ commutes (resp., anticommutes) 
with an even (resp., odd) supermatrix. 

The Hermitian conjugate is defined by  
\begin{eqnarray}
  {\cal B}^\dagger
  =
  \left(
    \begin{array}{cc}
      B_1^\dagger & F_2^\dagger\\
      F_1^\dagger & B_2^\dagger
    \end{array}
  \right),
  \qquad
  {\cal F}^\dagger
  =
  \left(
    \begin{array}{cc}
      F_1^{\prime\dagger} & B_2^{\prime\dagger}\\
      B_1^{\prime\dagger} & F_2^{\prime\dagger}
    \end{array}
  \right). 
\end{eqnarray}
A supermatrix ${\cal S}$ is {\it Hermitian} 
if ${\cal S}^\dagger={\cal S}$. 
Then, Hermitian matrices can be written as 
\begin{eqnarray}
  {\cal B}
  =
  \left(
    \begin{array}{cc}
      B_1 & F\\
      F^\dagger & B_2
    \end{array}
  \right), 
  \qquad
  {\cal F}
  =
  \left(
    \begin{array}{cc}
      F_1 & B\\
      B^\dagger & F_2
    \end{array}
  \right), 
  \qquad 
\end{eqnarray}
where $B_i$ and $F_i$ are bosonic and fermionic 
Hermitian matrices, and 
$B$ and $F$ are bosonic and fermionic complex matrices. 

The {\it supertrace} is defined by 
\begin{eqnarray}
  & &
  Str\ {\cal B}
  =
  Str
  \left(
    \begin{array}{cc}
      B_1 & F_1\\
      F_2 & B_2
    \end{array}
  \right)
  =
  tr\ B_1-tr\ B_2, 
  \\
  & &
  Str\ {\cal F}
  =
  Str
  \left(
    \begin{array}{cc}
      F_1^\prime & B_1^\prime\\
      B_2^\prime & F_2^\prime
    \end{array}
  \right)
  =
  tr\ F_1^\prime+tr\ F_2^\prime. 
\end{eqnarray}
It satisfies the cyclicity relation  
\begin{eqnarray}
  Str\left({\cal S}_1{\cal S}_2\right)
  =
  (-)^{s_1s_2}Str\left({\cal S}_2{\cal S}_1\right),    
\end{eqnarray}
where $s_i$ is $0$ (resp., $1$) if ${\cal S}_{i}$ is even (resp., odd). 
It also satisfies the following property for scalar supermatrices 
$\hat{c}$ and $\hat{\epsilon}$: 
\begin{eqnarray}
	Str\left(\hat{c}{\cal S}\right)
	=
	c\cdot Str\ {\cal S}, 
	\qquad
	Str\left(\hat{\epsilon}{\cal S}\right)
	=
	\epsilon\cdot Str\ {\cal S}. 
\end{eqnarray}

Finally, the {\it superdeterminant} is defined for even supermatrices by 
\begin{eqnarray}
  sdet {\cal B}
  =
  \exp\left(
    Str\log {\cal B}
  \right). 
\end{eqnarray}
\section{Derivation of Eqs. (\ref{eq:EOM_component_1}), 
(\ref{eq:EOM_component_2}), (\ref{eq:EOM_component_3}) 
and (\ref{eq:EOM_component_4})  
from Eqs.       
(\ref{eq:torsion+EOM1}) and (\ref{eq:torsion+EOM5})}
\label{sec:Derivation of EOM}
\setcounter{equation}{0} 
In this appendix we show that 
if the torsion constraints 
and equations of motion of the four-dimensional supergravity 
in superspace, Eqs. (\ref{eq:torsion+EOM1}) and (\ref{eq:torsion+EOM5}), 
are satisfied, then Eqs. 
(\ref{eq:EOM_component_1}), (\ref{eq:EOM_component_2}) 
and (\ref{eq:EOM_component_3}), and hence the equations of motion 
of the matrix model, are satisfied.  
The argument is almost parallel to that given in Ref. \citen{WZ77}.  
\subsection{Bianchi identities}
\label{subsec:Bianchi identities}
The supertorsion and the super Riemann tensor satisfy 
the following Bianchi identities which arise from the 
super Jacobi identities. If we use Eqs. 
(\ref{eq:torsion+EOM1}) and (\ref{eq:torsion+EOM5}) to 
simplify the Bianchi identities,  
\\
$[{\nabla}_a,[{\nabla}_b,{\nabla}_c]]+({\rm cyclic})=0$ gives
\begin{eqnarray}
  & &
  R_{[abc]}{}^{i}
  =
  0, 
  \label{eq:Bianchi_1}
  \\
  & &
  {\nabla}_{[a} R_{bc]}{}^{ij}
  +
  T_{[ab}{}^\gamma R_{c]\gamma}{}^{ij}
  =
  0, 
  \label{eq:Bianchi_2}
\end{eqnarray}
$[{\nabla}_a,[{\nabla}_b,{\nabla}_\gamma]]+({\rm cyclic})=0$ gives 
\begin{eqnarray}
  & &
  -R_{b\gamma a}{}^i
  + 
  R_{a\gamma b}{}^i
  -
  2iT_{ab}{}^\delta 
  \left(\gamma^i\right)_{\gamma\delta}
  =
  0, 
  \label{eq:Bianchi_4}
  \\
  & &
  \frac{1}{4}R_{ab}{}^{ij}
  \left(\gamma_{ij}\right)_\gamma{}^\delta
  -
  {\nabla}_\gamma T_{ab}{}^\delta
  =
  0,
  \label{eq:Bianchi_6}
\end{eqnarray}
and 
$[{\nabla}_a,\{{\nabla}_\beta,{\nabla}_\gamma\}]
+\{{\nabla}_\beta,[{\nabla}_\gamma,{\nabla}_a]\}
-\{{\nabla}_\gamma,[{\nabla}_a,{\nabla}_\beta]\}=0$ 
gives 
\begin{eqnarray}
  & &
  R_{\beta\gamma}{}^{ij}=0, 
  \label{eq:Bianchi_7}
  \\
  & &
  2i\left(\gamma^k\right)_{\beta\gamma}
  R_{ak}{}^{ij}
  -
  {\nabla}_\beta R_{a\gamma}{}^{ij}
  -
  {\nabla}_\gamma R_{a\beta}{}^{ij}
  =
  0, 
  \label{eq:Bianchi_8}
  \\
  & &
  2i\left(\gamma^k\right)_{\beta\gamma}
  T_{ak}{}^\delta
  -
  \frac{1}{4}R_{a\gamma}{}^{ij}
  \left(\gamma_{ij}\right)_\beta{}^\delta
  -
  \frac{1}{4}R_{a\beta}{}^{ij}
  \left(\gamma_{ij}\right)_\gamma{}^\delta
  = 
  0, 
  \label{eq:Bianchi_9}
\end{eqnarray}
where we have used Eq. (\ref{eq:Bianchi_7}) to simplify 
Eq. (\ref{eq:Bianchi_8}). 
Here we have presented only the identities that we need  
in the following. 
\subsection{Derivation of Eqs. (\ref{eq:EOM_component_1}), 
(\ref{eq:EOM_component_2}) and (\ref{eq:EOM_component_3}) }
\subsubsection{Proof of $R_{ab}{}^{cd}=R^{cd}{}_{ab}$}%
From the fact that $R_{ab}{}^{cd}$ is antisymmetric 
under the exchanges $a\leftrightarrow b$ and $c\leftrightarrow d$  
and Eq. (\ref{eq:Bianchi_1}), it follows that 
\begin{eqnarray}
  R_{abcd}
  &=&
  -R_{bcad}-R_{cabd}
  \nonumber\\
  &=&
  -R_{cbda}+R_{cadb}
  \nonumber\\
  &=&
  R_{dcba}+R_{bdca}
  -R_{adcb}-R_{dcab}
  \nonumber\\
  &=&
  2R_{cdab}+R_{dbac}+R_{adbc}
  \nonumber\\
  &=&
  2R_{cdab}-R_{badc}
  \nonumber\\
  &=&
  2R_{cdab}-R_{abcd}. 
\end{eqnarray}
Hence, we have 
\begin{eqnarray}
  R_{ab}{}^{cd}=R^{cd}{}_{ab}. 
  \label{symmetry of Riemann tensor}
\end{eqnarray}
\subsubsection{Proof of the Rarita-Schwinger equation}
Multiplying Eq. (\ref{eq:Bianchi_9}) by 
$\frac{1}{4}(\gamma_b)^{\gamma\beta}$, we have 
\begin{eqnarray}
  2iT_{ab}{}^\delta
  &=&
  \frac{1}{8}R_{a\beta}{}^{ij}(\gamma_b\gamma_{ij})^{\beta\delta}
  \nonumber\\
  &=&
  \frac{1}{8}R_{a\beta}{}^{ij}
  \left(
    \eta_{bi}\gamma_j
    -
    \eta_{bj}\gamma_i
    +
    \gamma_{bij}
  \right)^{\beta\delta}
  \nonumber\\
  &=&  
  \frac{1}{4}R_{a\beta b}{}^{j}(\gamma_j)^{\beta\delta}
  +
  \frac{1}{8}R_{a\beta}{}^{ij}(\gamma_{bij})^{\beta\delta}
  \nonumber\\
  &=&  
  \frac{1}{4}R_{a\beta b}{}^{j}(\gamma_j)^{\delta\beta}
  -
  \frac{1}{8}R_{a\beta}{}^{ij}(\gamma_{bij})^{\delta\beta}. 
\end{eqnarray}
Then, multiplying this by $(\gamma^b)^\gamma{}_\delta$, we obtain 
\begin{eqnarray}
  2i(\gamma^b)^\gamma{}_\delta T_{ab}{}^\delta
  &=&
  \frac{1}{4}R_{a\beta}{}^{bj}(\gamma_{bj})^{\gamma\beta}
  -
  \frac{1}{8}R_{a\beta}{}^{ij}(\gamma^b\gamma_{bij})^{\gamma\beta}
  \nonumber\\
  &=&
  \frac{1}{4}R_{a\beta}{}^{ij}(\gamma_{ij})^{\gamma\beta}
  -
  \frac{4-2}{8}R_{a\beta}{}^{ij}(\gamma_{ij})^{\gamma\beta}
  \nonumber\\
  &=&
  0.  
\end{eqnarray}
On the other hand, by contracting $\gamma$ and $\delta$ in 
(\ref{eq:Bianchi_9}), we have  
\begin{eqnarray}
  2i(\gamma^b)^\gamma{}_\delta T_{ab}{}^\delta
  =
  \frac{1}{4}R_{a\beta}{}^{ij}(\gamma_{ij})^{\gamma\beta}. 
\end{eqnarray}
Therefore, we find 
\begin{eqnarray}
  (\gamma^a)_{\alpha\beta}T_{ab}{}^\beta
  =
  0, 
  \qquad
  R_{a\beta}{}^{ij}(\gamma_{ij})^{\gamma\beta}
  =
  0. 
  \label{eq:RS}
\end{eqnarray}
The former is the Rarita-Schwinger equation 
written in terms of superspace \cite{WZ77}. 
\subsubsection{Proof of the Einstein equation (\ref{eq:EOM_component_1})} 
By contracting $a$ and $i$ in Eq. (\ref{eq:Bianchi_4}) and 
using the second equation in Eq. (\ref{eq:RS}), 
we find that 
\begin{eqnarray}
  R_{a\gamma}{}^{ac}=0. 
\end{eqnarray}
Next, contracting $a$ and $i$ in Eq. (\ref{eq:Bianchi_8}) and 
substituting the above equation, we obtain 
\begin{eqnarray}
  (\gamma^k)_{\beta\gamma}R_{ak}{}^{aj}=0,  
\end{eqnarray}
which is equivalent to Eq. (\ref{eq:EOM_component_1}):
\begin{eqnarray}
  R_{ab}{}^{ac}=0.\label{eq:Einstein} 
\end{eqnarray}
This is the Einstein equation without a cosmological constant. 
Note that the super Ricci tensor $R_{ab}{}^{ac}$ contains 
the contribution from the energy-momentum of the gravitino. 
\subsubsection{Proof of Eq. (\ref{eq:EOM_component_2})}
By Eqs. (\ref{eq:torsion+EOM1}) and (\ref{eq:Bianchi_7}), we have 
\begin{eqnarray}
  \{{\nabla}_\alpha,{\nabla}_\beta\}
  &=&
  2i(\gamma^a C^{-1})_{\alpha\beta}{\nabla}_a\ , 
  \\
  {\nabla}_a
  &=&
  -\frac{i}{8}\left(C\gamma_a\right)^{\alpha\beta}
  \{{\nabla}_\alpha,{\nabla}_\beta\}
  \nonumber\\
  &=&
  -\frac{i}{8}\left(C\gamma_a\right)^{\alpha\beta}
  {\nabla}_\alpha{\nabla}_\beta\ . 
  \label{nabla_a_and_nabla_alpha}
\end{eqnarray}
Then, multiplying Eq. (\ref{eq:Bianchi_6}) by 
$(\gamma^a)_\rho{}^\gamma$, we have 
\begin{eqnarray}
  (\gamma^a)_\rho{}^\gamma{\nabla}_\gamma T_{ab}{}^\delta
  &=&
  \frac{1}{4}(\gamma^a\gamma^{ij})_\rho{}^\delta
  R_{abij}
  \nonumber\\
  &=&
  \frac{1}{4}
  \left(
    \gamma^{aij}
    +
    \eta^{ai}\gamma^j
    - 
    \eta^{aj}\gamma^i
  \right)_\rho{}^\delta
  R_{abij}
  \nonumber\\
  &=&
  0. 
  \label{eq:RS2}
\end{eqnarray}
Here we have used the relation $R_{a[bcd]}=0$, which follows from 
Eqs. (\ref{symmetry of Riemann tensor}) and (\ref{eq:Bianchi_1}), 
and Eq. (\ref{eq:Einstein}). 
Next, multiplying Eq. (\ref{eq:RS2}) by ${\nabla}^\rho$ and using 
Eq. (\ref{nabla_a_and_nabla_alpha}), 
we obtain Eq. (\ref{eq:EOM_component_2}): 
\begin{eqnarray}
  {\nabla}^a T_{ab}{}^\delta=0. 
\end{eqnarray}  
\subsubsection{Proof of Eqs. (\ref{eq:EOM_component_3}) and 
  (\ref{eq:EOM_component_4}) }
From Eqs. (\ref{eq:RS}), (\ref{eq:Bianchi_4}) and
(\ref{eq:Bianchi_9}),  
it follows that \cite{WZ77,WB}
\begin{eqnarray}
  R_{\alpha b}{}^{cd}
  =
  2i(\gamma_b C^{-1})_{\alpha\beta}T^{cd\beta}\label{eq:R alpha bcd}. 
\end{eqnarray}
Then, because $C\gamma^a$ is symmetric, we have  
\begin{eqnarray}
  T_{bc}{}^\gamma R^a{}_{\gamma}{}^{ij}=
  T^{ij\gamma} R^a{}_{\gamma}{}_{bc}. 
\end{eqnarray}
Combining this with Eq. (\ref{symmetry of Riemann tensor}), 
we can rewrite Eq. (\ref{eq:Bianchi_2}) as 
\begin{eqnarray}
  {\nabla}^{[a}R_{ij}{}^{bc]}
  +
  T_{ij}{}^\gamma R^{[a}{}_\gamma{}^{bc]}
  =
  0. 
\end{eqnarray}
The second term on the right-hand side is zero, because substituting 
Eq. (\ref{eq:R alpha bcd}) into Eq. (\ref{eq:Bianchi_4}) we have 
$R^{[a}{}_\gamma{}^{bc]}=0$. 
Therefore, we obtain 
\begin{eqnarray}
  {\nabla}^{[a}R_{ij}{}^{bc]}=0, 
\end{eqnarray}
and contracting $a$ and $i$, we have 
\begin{eqnarray}
  {\nabla}^a R_{ab}{}^{cd}
  +
  {\nabla}^c R_{ab}{}^{da}
  +
  {\nabla}^d R_{ab}{}^{ac}
  =
  0. 
\end{eqnarray}
Because the second and third terms are zero, by Eq. (\ref{eq:Einstein}), 
we obtain Eq. (\ref{eq:EOM_component_3}): 
\begin{eqnarray}
  {\nabla}^a R_{ab}{}^{cd}=0\ . 
\end{eqnarray}

From Eq. (\ref{eq:R alpha bcd}) we have
\begin{eqnarray}
  T_{ab}{}^\gamma R^a{}_\gamma{}^{ij}
  &=&
  T_{ab}{}^\gamma\cdot
  (-2i)(\gamma^a C^{-1})_{\gamma\delta}T^{ij\delta}
  \nonumber\\
  &=&
  2iT^{ij\delta}\cdot
  (\gamma^a C^{-1})_{\delta\gamma}T_{ab}{}^\gamma, 
\end{eqnarray}
but this is zero, by Eq. (\ref{eq:RS}). 
Therefore, we have shown Eq. (\ref{eq:EOM_component_4}). 
\section{Proof of $\kappa=0$}
\label{subsec:kappa=0}
\setcounter{equation}{0} 
In the footnote of 
\S \ref{subsec:Classical Solutions_super}, we claimed 
that if we stipulate  
${\cal A}_a=i{\nabla}_{(a)},\ \Psi_\alpha=\kappa{\nabla}_{(\alpha)}$ 
and the standard off-shell torsion constraints 
(\ref{eq:torsion+EOM1}) as an ansatz, 
then the equations of motion (\ref{eq:EOM1}) and (\ref{eq:EOM2}) 
of the matrix model force $\kappa$ to be zero. 
In this appendix, we give an outline of its proof. 

By combining Eq. (\ref{eq:torsion+EOM1}) 
with the Bianchi identities, 
the remaining degrees of freedoms can be expressed 
in terms of three superfields: the complex scalar $R$, 
the real vector $G_a$, and the spin $\frac{3}{2}$ field 
$W_{(\alpha\beta\gamma)}$ \cite{GWZ79,WB}. 

If $\kappa\neq 0$,  
Eq. (\ref{eq:EOM2}) reduces to  
\begin{eqnarray}
  (\gamma^a)_\alpha{}^\beta
  \left(
    R_{a\beta}{}^{cd}{\cal O}_{cd}
    +
    T_{a\beta}{}^\gamma{\nabla}_\gamma
  \right)=0, 
\end{eqnarray}
which is equivalent to 
\begin{eqnarray}
  (\gamma^a)_\alpha{}^\beta R_{a\beta}{}^{cd}
  &=&
  0, 
  \label{W=0}
  \\
  (\gamma^a)_\alpha{}^\beta T_{a\beta}{}^\gamma
  &=&
  0. 
  \label{G=R=0}
\end{eqnarray}
In terms of $G,R$ and $W$, the latter becomes \footnote{
The formulae in Chapter XV of Ref. \citen{WB} are useful. 
}
\begin{eqnarray}
  G_a=R=0. 
\end{eqnarray}
This is the equation of motion of supergravity \cite{GWZ78}, 
and it is equivalent to Eq. (\ref{eq:torsion+EOM5}). 
As explained in the previous subsection, 
this implies Eq. (\ref{eq:Einstein}); that is, 
the cosmological constant is zero. 
(Furthermore, Eq. (\ref{W=0}) gives $W_{(\alpha\beta\gamma)}=0$, 
which with Eq. (\ref{G=R=0}) allows only a flat spacetime.)
On the other hand,  
substituting Eqs. (\ref{ansatz:matrix and covariant derivative}) 
and (\ref{eq:torsion+EOM1}) 
into Eq. (\ref{eq:EOM1}), the coefficient of ${\nabla}_c$ becomes 
\begin{eqnarray}
  i\left(
    R_{ab}{}^{ac}+8\kappa^2\delta_b{}^c
  \right)
  =
  0. 
\end{eqnarray}
Hence we find that the cosmological constant is given by 
$\Lambda=-32\kappa^2$. 
This is a contradiction, and thus $\kappa$ must be zero.


\begin{thebibliography}{99}
\bibitem{HKK}
  M. Hanada, H. Kawai and Y. Kimura, 
  Prog. Theor. Phys. \textbf{114} (2005), 1295;  
  hep-th/0508211. 
  
\bibitem{BFSS}
  T. Banks, W. Fischler, S. Shenker and L. Susskind, 
  Phys. Rev. D \textbf{55} (1997), 5112; hep-th/9610043.
  
\bibitem{IKKT}
  N. Ishibashi, H. Kawai, Y. Kitazawa and A. Tsuchiya, 
  Nucl. Phys. B \textbf{498} (1997), 467; hep-th/9612115.
  
\bibitem{EK}
  T. Eguchi and  H. Kawai ,
  Phys. Rev. Lett. \textbf{48}(1982), 1063. 
  
\bibitem{hep-th/0204078}
  T. Azuma and H. Kawai, 
  Phys. Lett. B \textbf{538} (2002), 393; 
  hep-th/0204078. 
  
\bibitem{eigenvalue}
  H. Aoki, S. Iso, H. Kawai, Y. Kitazawa and T. Tada, 
  Prog. Theor. Phys. \textbf{99} (1998), 713;  
  hep-th/9802085; 
  J. Nishimura and F. Sugino,
  J. High Energy Phys. \textbf{05} (2002), 001; 
  hep-th/0111102. 

  H. Kawai, S. Kawamoto, T. Kuroki, T. Matsuo and S. Shinohara,
  Nucl. Phys. B \textbf{647} (2002), 153;  
  hep-th/0204240. 

  H. Kawai, S. Kawamoto, T. Kuroki and S. Shinohara,
  Prog. Theor. Phys. \textbf{109} (2003), 115;  
  hep-th/0211272.

\bibitem{Howe}
  P. S. Howe, 
  J. of Phys. A \textbf{12} (1979), 393. 
 
  M. F. Ertl, 
  hep-th/0102140. 
  
\bibitem{WZ77}
  J. Wess and B. Zumino, 
  Phys. Lett. B \textbf{66} (1977), 361. 

\bibitem{WB}
  J. Wess and J. Bagger,
  {\it Supersymmetry and Supergravity}, second edition ( 
  Princeton Univ. Press, 1992). 
  
\bibitem{GWZ78}
  R. Grimm, J. Wess and B. Zumino, 
  Phys. Lett. B \textbf{73} (1978), 415. 
 
  J. Wess and B. Zumino, 
  Phys. Lett. B \textbf{74} (1978), 51. 
  
\bibitem{WZ78Dec}
  J. Wess and B. Zumino, 
  Phys. Lett. B \textbf{79} (1978), 4. 
  
\bibitem{GWZ79}
  R. Grimm, J. Wess and B. Zumino, 
  Phys. Lett. B \textbf{73} (1978), 415. 

\bibitem{hep-th/9612222}
  M. Li, 
  Nucl. Phys. B \textbf{499} (1997), 149; hep-th/9612222. 
 
  A. Connes, M. R. Douglas and A. Schwarz, 
  J. High Energy Phys. \textbf{02} (1998), 003; 
  hep-th/9711162. 
 
  H. Aoki, N. Ishibashi, S. Iso, H. Kawai, Y. Kitazawa and T. Tada, 
  Nucl. Phys. B \textbf{565} (2000), 176;  
  hep-th/9908141. 

\bibitem{ADR}
  J. Atick, A. Dhar and B. Ratra, 
  Phys. Rev. D \textbf{33} (1986), 2824. 
  
  B. E. W. Nilsson and A. K. Tollsten, 
  Phys. Lett. B \textbf{171} (1986), 212. 

\bibitem{Alvarez-Gaume}  
  L. Alvarez-Gaume and J. L. Manes, 
  Mod. Phys. Lett. A \textbf{6} (1991), 2039. 
  S. A. Yost, 
  Int. J. Mod. Phys. A \textbf{7} (1992), 6105; hep-th/9111033. 

\bibitem{OT}
  T. Okuda and T. Takayanagi, 
  J. High Energy Phys. \textbf{03} (2006), 062; 
  hep-th/0601024. 

\bibitem{DeWitt}
  B. DeWitt, 
  {\it Supermanifolds}, second edition 
  (Cambridge Univ. Press, 1992). 
\end{thebibliography}
\end{document}